# A Qualitative Study of User Perception of M365 AI Copilot

**Muneera Bano, Didar Zowghi, Jon Whittle, Liming Zhu, Andrew Reeson**
firstname.lastname@csiro.au
*CSIRO's Data61, Australia*

**Rob Martin**
firstname.lastname@csiro.au
*IM&T, CSIRO, Australia*

**Jen Parsons**
firstname.lastname@csiro.au
*Digital Office, CSIRO, Australia*

## Abstract

The adoption of AI copilots in professional workflows presents opportunities for enhanced productivity, efficiency, and decision-making. In this paper we present results from a six-month trial of M365 Copilot that was conducted at our organisation during 2024. A qualitative interview study was conducted with 27 participants. The study explored user perceptions of M365 Copilot's effectiveness, productivity impact, evolving expectations, ethical concerns, and overall satisfaction. Initial enthusiasm for the tool was met with mixed post-trial experiences. While some users found M365 Copilot beneficial for tasks such as email coaching, meeting summaries, and content retrieval, some reported unmet expectations in areas requiring deeper contextual understanding, reasoning, and integration with existing workflows. Ethical concerns were a recurring theme, with users highlighting issues related to data privacy, transparency, and AI bias. While M365 Copilot demonstrated value in specific operational areas, its broader impact remained constrained by usability limitations and the need for human oversight to validate AI-generated outputs.

## 1. Introduction

The integration of AI copilots into professional environments is rapidly reshaping workplace productivity, efficiency, and decision-making. Microsoft 365 (M365)'s AI-powered assistant embedded within applications such as Microsoft 365, GitHub, and Bing, represents a shift in how artificial intelligence is leveraged for knowledge work (Kytö 2024). Designed to automate routine tasks, streamline workflows, and provide contextual recommendations, Copilot holds promise for improving organisational productivity and efficiency (Dell'Acqua et al. 2023; Moroz, Grizkevich, and Novozhilov 2022; Smit et al. 2024; Sun, Che, and Wang 2024; Ziegler et al. 2024). However, as AI copilots become increasingly embedded in workflows, the question remains on their impact on individual and organisation wide productivity.

This study examines the trial of M365 Copilot within Commonwealth Scientific and Industrial Research Organisation (CSIRO)[1], exploring user perceptions and experiences of its impact on productivity, efficiency, effectiveness, ethical concerns, and evolving user expectations and satisfaction. AI copilots are marketed[2] as transformative technologies, but their real-world value depends on several socio-technical factors, including integration with existing workflows, user trust, and the extent to which they align with professional demands.

---

[1] https://www.csiro.au/en/

[2] https://news.microsoft.com/reinventing-productivity/

The research presented in this paper forms part of a larger study on M365 Copilot trial at CSIRO. A broader quantitative survey study (Bano, Zowghi, et al. 2024) was conducted to assess the tool's overall impact on productivity, efficiency, ethical concerns and user satisfaction from 300 participants of CSIRO. The survey results indicated mixed outcomes, with participants reporting improvements in structured tasks but also identifying limitations in Copilot's advanced functionalities and integration. To complement the survey findings, in-depth qualitative interviews were conducted with 27 participants sharing their experiences.

Findings from this study reveal that while M365 Copilot provided measurable improvements in certain areas, such as meeting summarisation, email drafting, and basic information retrieval, it fell short in areas requiring domain-specific knowledge, creative problem-solving, and nuanced decision-making. Users reported a productivity paradox, where time saved through automation was often offset by the need for extensive verification and correction of AI-generated outputs. Ethical concerns—including data privacy, bias, and transparency—also influenced user trust and adoption levels, particularly in knowledge-intensive roles.

As organisations weigh the return on investment for AI copilots, they must consider whether these tools genuinely enhance productivity or simply shift cognitive effort elsewhere. This study provides insights for researchers, practitioners, and policymakers, offering a grounded evaluation of AI copilots in high-stakes professional environments. By examining the intersection of productivity, efficiency, ethical concerns, and evolving expectations, the research highlights key considerations for the responsible and effective integration of AI copilots in modern workplaces.

This paper is structured as follows. Section 2 reviews related literature on AI copilots and their impact on productivity, efficiency, and ethics. Section 3 outlines the research methodology, including participant selection and data analysis. Section 4 presents the study's findings, while Section 5 discusses key themes such as usability, ethical concerns, and evolving user expectations. Section 6 addresses study limitations, and Section 7 concludes and recommendations for AI copilot adoption.

## 2. Background

The integration of artificial intelligence (AI) into professional settings is increasingly reshaping workflows, with AI copilots emerging as a critical development in knowledge-intensive domains (Coyle and Jeske 2023; Sellen and Horvitz 2024). AI copilots, such as M365 Copilot[3], Google's Duet AI[4] and Anthropic's Claud 2[5], are designed to support users by automating routine tasks, providing real-time suggestions, and facilitating decision-making. These systems leverage large language models (LLMs) to interpret user inputs and generate contextually relevant outputs, thereby potentially enhancing efficiency and productivity (Peng et al. 2023; Bano, Hoda, et al. 2024). However, while AI copilots promise to augment human capabilities, their actual impact varies significantly across domains and remains an area of active investigation.

The concept of AI Copilots has roots in aviation, where the idea of an "automated Copilot" began to support pilots with complex decision-making and reduce their cognitive load. Early AI Copilots in aviation, like HARVIS (Bejarano et al. 2022) and V-CoP (Li et al. 2024), were designed to aid single-pilot operations by offering quick-access procedures, dynamic rerouting, and real-time situational analysis. In contrast to traditional automated systems, these Copilots introduced a more collaborative relationship between human pilots and technology, setting the foundation for AI Copilots that extend beyond aviation into diverse professional domains.

It is important to differentiate between AI Copilots and other AI-driven systems, such as *AI assistants*, *conversational agents*, or *autonomous systems* (Hayawi and Shahriar 2024). While AI assistants like Siri or

---

[3] https://blogs.microsoft.com/blog/2023/09/21/announcing-microsoft-Copilot-your-everyday-ai-companion/
[4] https://workspace.google.com/blog/product-announcements/duet-ai
[5] https://www.anthropic.com/news/claude-2

Google Assistant respond to user commands and perform specific functions, they are often limited to single-use tasks and require ongoing user input. AI Copilots, in contrast, work alongside users, engaging in tasks collaboratively by providing real-time suggestions, recommendations, and intelligent insights that adapt dynamically to changing contexts. Unlike AI agents or autonomous systems that can operate independently to achieve goals—like virtual customer service agents or robotic process automation systems—AI Copilots enhance user control and decision-making, offering a partnership model rather than full autonomy. In this sense, Copilots sits at a unique intersection: they empower users without diminishing their control, making complex tasks more manageable while preserving human agency.

As AI capabilities advanced, the notion of a Copilot evolved, moving from aviation into fields like software development (Moroz, Grizkevich, and Novozhilov 2022; Russo 2024; Smit et al. 2024), healthcare (Menz et al. 2024; Rossettini et al. 2024), education (Adetayo, Aborisade, and Sanni 2024; Tepe and Emekli 2024; Wermelinger 2023; Puryear and Sprint 2022), and business intelligence (Sellen and Horvitz 2024; Coyle and Jeske 2023; Dell'Acqua et al. 2023). GitHub Copilot[6], launched in 2021, brought the concept of an AI-powered coding assistant to the programming community, enabling developers to generate code suggestions, solve algorithmic challenges, and automate repetitive tasks (Dakhel et al. 2023; Peng et al. 2023; Ziegler et al. 2024). Its success highlighted the potential of AI Copilots in enhancing productivity across professional settings, and it paved the way for M365 Copilot, a tool integrated into Microsoft 365, designed to support knowledge workers with tasks such as drafting, summarising, and data analysis.

The application of AI copilots has been studied in various fields, yielding mixed results in terms of productivity gains and efficiency improvements. In software development, GitHub Copilot has been associated with time savings in coding tasks (Peng et al. 2023), while in marketing analytics, copilots have helped streamline campaign optimisation and data-driven decision-making (Coyle and Jeske 2023). Knowledge workers, including consultants and writers, report benefits from the automation of routine tasks, enabling greater focus on strategic and creative work (Dell'Acqua et al. 2023; Ziegler et al. 2024). In research contexts, LLM-based copilots such as ChatGPT-3.5 and GPT-4 have supported human reasoning and classification without replacing analysts (Bano, Hoda, et al. 2024; Bano, Zowghi, and Whittle 2023). Despite these advances, the literature underscores the ongoing need for rigorous evaluation and robust safeguards, particularly in high-stakes domains like healthcare and cybersecurity, where issues of accuracy, privacy, and ethical deployment remain critical (Hannon et al. 2024; Menz et al. 2024).

*Measuring Productivity, Efficiency, and Effectiveness in AI Copilots*

The concepts of productivity, efficiency, and effectiveness serve as fundamental benchmarks for evaluating AI copilots, yet their definitions and applications vary widely across disciplines. Productivity generally refers to the quantity of outputs produced relative to inputs, such as the number of completed tasks or research publications (Pritchard 1995). Efficiency, in contrast, focuses on resource optimisation, including time, effort, and cost reductions to achieve desired outputs with minimal waste (Lovell 1993). Effectiveness, meanwhile, prioritises the quality and impact of outcomes, assessing whether objectives are met satisfactorily (Neely, Gregory, and Platts 1995). These distinctions are particularly important in AI research, where copilots must balance high output, resource optimisation, and meaningful goal alignment.

In AI-assisted workflows, productivity is often shaped by domain-specific demands. In software development, productivity is measured in terms of coding efficiency, debugging time, and feature implementation, with AI copilots such as GitHub Copilot helping automate routine coding tasks and allowing developers to focus on complex problem-solving (Bird et al. 2022; Dakhel et al. 2023). In marketing, AI copilots facilitate productivity by optimising campaigns, generating content, and deriving actionable insights from data (Coyle and Jeske 2023)). In research and knowledge work, productivity is linked to information synthesis, report generation, and decision-making efficiency, where LLM-powered copilots reduce cognitive load and automate repetitive tasks (Sellen and Horvitz 2024).

The inconsistencies in defining productivity, efficiency, and effectiveness across AI Copilot studies stem from several key factors. One major factor is domain-specific objectives and metrics, as AI Copilots serve distinct functions across fields such as software development, healthcare, education, security, and business intelligence. Additionally, the evolving functionalities of AI Copilots contribute to these inconsistencies. Earlier iterations primarily focused on automating routine tasks, which made them well-suited to productivity and efficiency metrics. However, newer models like ChatGPT-4 and Claude 2 exhibit advanced contextual reasoning, shifting the emphasis towards effectiveness, particularly in complex domains that require interpretive and advisory capabilities (Coyle and Jeske 2023; Menz et al. 2024). As AI Copilots continue to evolve, expectations regarding their impact on work processes will likely expand, further influencing how productivity, efficiency, and effectiveness are measured.

---

[6] https://github.com/features/copilot

Methodological variability and cross-disciplinary interpretations also contribute to these inconsistencies. Different studies prioritise different aspects of Copilot performance, with some focusing on quantifiable task completion time to measure productivity, while others assess qualitative aspects such as output reliability. In education, productivity might be linked to student engagement metrics, whereas in cybersecurity, effectiveness is more critical due to the need for precise threat detection (Chen 2024; Hannon et al. 2024). Furthermore, the interpretation of these terms varies across disciplines—computer science typically quantifies productivity and efficiency, while healthcare and education assess effectiveness in terms of clinical outcomes or learning improvements. As AI Copilots become more embedded in professional workflows, standardising definitions and evaluation metrics will be crucial for enabling meaningful comparisons of Copilot impact across different domains.

### *Ethical and Security Considerations*

Although AI copilots claimed to offer substantial productivity benefits, they also introduce ethical concerns and security risks that warrant close examination. Research on GitHub Copilot, for example, has revealed that approximately 40% of its AI-generated code contains security vulnerabilities, raising questions about its reliability in high-risk software development environments (Pearce et al. 2022). More broadly, AI copilots raise ethical issues concerning privacy, data governance, and intellectual property. As these systems rely on extensive training datasets—often including proprietary or sensitive information—there are growing concerns about data security and potential privacy breaches (Coyle and Jeske 2023). Additionally, questions regarding the intellectual ownership of AI-generated content remain unresolved, particularly in creative and technical fields (Moroz, Grizkevich, and Novozhilov 2022).

Another emerging issue is the potential for de-skilling and automation bias ((Nazareno and Schiff 2021). As AI copilots increasingly handle routine tasks, there is a risk that human users may become overly dependent on them, leading to a decline in fundamental skills and critical thinking. This concern is particularly pronounced in software development, where maintaining a deep understanding of coding principles is essential for ensuring quality and innovation (Sellen and Horvitz 2024). Furthermore, AI-generated outputs can contribute to automation bias, wherein users accept AI-generated recommendations uncritically, potentially exacerbating errors or reinforcing existing biases (Horne 2023).

### *AI Copilots in Scientific Research*

Despite their widespread adoption, AI Copilots remain underexplored in scientific research, where their potential to enhance research productivity, streamline processes, and assist in data analysis is still largely unknown (Bano, Hoda, et al. 2024; Bano et al. 2023; Bano, Zowghi, and Whittle 2023). This gap presents an opportunity for further studies to examine how AI Copilots can specifically contribute to scientific discovery and innovation.

AI Copilots hold immense potential to transform scientific research, though domain-specific evidence supporting their integration remains limited (Bano, Hoda, et al. 2024). Globally, countries are recognising the transformative power of AI and are investing in strategies to leverage its potential while addressing ethical concerns. For example, Australia's strategic initiatives highlight AI's ability to drive innovation across industries such as healthcare, agriculture, and manufacturing, emphasising principles of data privacy, accountability, and transparency (Hajkowicz et al. 2023). Similarly, the United States' National Artificial Intelligence Research and Development Strategic Plan highlights AI's role in advancing scientific innovation, though it stops short of explicitly addressing AI Copilots in research workflows ('Office of Science and Technology Policy (2023), The National Artificial Intelligence Research and Development Strategic Plan.' 2023). However, explicit discussions about AI Copilots or generative AI systems in the context of scientific research remain rare. One study (Toner-Rodgers 2024) has provided understandings into the impact of AI in research contexts. Their findings demonstrate that AI-assisted scientists achieve substantial productivity gains, including a 44% increase in material discoveries, a 39% rise in patent filings, and a 17% boost in product prototypes. However, disparities persist, as lower-performing researchers benefit less and express concerns about reduced creativity and job satisfaction, underscoring the need for AI tools to complement domain expertise effectively (Bano, Zowghi, and Whittle 2023).

Our literature review has identified gaps in ethical awareness within AI applications in science, further emphasising the need for robust strategies to ensure the responsible adoption of AI in research settings (Bano et al. 2023). Collectively, these global efforts illustrate the growing recognition of AI's potential in scientific domains but also highlight the need for focused exploration of AI Copilots' unique contributions to productivity, efficiency, and ethical considerations in science and research.

## 3. Research Methodology

This study employed a qualitative research methodology to investigate user perceptions of M365 Copilot within CSIRO, a scientific research institution. Through in-depth interviews, the study explored how users interpreted productivity, efficiency, and effectiveness in their respective roles, along with their evolving expectations, ethical concerns, and overall satisfaction with AI-assisted workflows.

### Research Objective

The primary objective of this interview study was to gain in-depth qualitative understandings into user experiences with M365 Copilot, focusing on how users perceived and understood their productivity, efficiency, and effectiveness within the Organisation. Unlike corporate or administrative settings, where AI-assisted tools may primarily support routine or transactional tasks, the integration of AI solutions in scientific research presents distinct challenges and opportunities. The study aimed to explore how Copilot influenced users' workflows, whether it streamlined tasks or introduced new complexities, and how ethical concerns related to data privacy, AI bias, and automation were perceived within a research-intensive context. By investigating these aspects, the research sought to provide a nuanced understanding of AI adoption in a setting where accuracy, analytical depth, and intellectual contribution are paramount.

### Participant Selection Process

To ensure diverse perspectives, a systematic persona-based selection approach was used to recruit participants from various professional roles, seniority levels, and demographic backgrounds. The study initially invited 40 participants from Organisation's broader M365 Copilot trial cohort, of which 27 agreed to participate in in-depth interviews. These individuals were drawn from a larger pool of 300 trial users, ensuring representation from different functional areas and levels of experience with AI-assisted workflows. The participants included researchers, research managers, operational support staff, and individuals involved in strategic decision-making, providing a comprehensive view of how Copilot was perceived and utilised across different roles within the organisation.

All user-reported data in this study was fully anonymised in accordance with the ethical guidelines set by Organisation's Human Ethics Research Committee. This measure was taken to ensure that participants could provide candid feedback without concerns about potential repercussions on their trial participation or access to M365 Copilot. Participation in the study was entirely voluntary, with explicit assurances that involvement in the research would have no impact on their access to the trial or their professional standing. All data collected was self-reported and obtained with informed consent, maintaining strict confidentiality and respecting the participants' professional environment.

### Interview Timeline

The interviews were conducted over Microsoft Teams around mid-2024, with each session lasting approximately one hour. This timing allowed participants sufficient time to familiarise themselves with Copilot following the trial's commencement in January 2024. A structured questionnaire, comprising 32 open-ended questions (see Appendix A), guided the discussions. The questions were designed to capture user perceptions of productivity, efficiency, and effectiveness, as well as evolving expectations, ethical concerns, and overall satisfaction with Copilot's capabilities. Given the complexity of AI integration in a research environment, the study focused on how users interpreted these concepts in their respective domains, rather than imposing predefined metrics of productivity and efficiency.

To maintain participant anonymity and avoid potential bias, all interviews were conducted by staff from the organisation's IT department, who were also responsible for the successful implementation of the trial and the ongoing management of trial participants. The IT staff were briefed to adhered to a standardised interview protocol developed by the research team. All interview were recorded and transcripts were anonymised before being shared with the research team for analysis, with identifying details removed to ensure confidentiality.

### Demographics of Interview Participants

The study cohort represented a balanced demographic distribution, with 13 women, 12 men, one non-binary participant, and one unspecified. Many participants had substantial experience in their fields, with over half having worked at the organisation for more than a decade. Additionally, 56% had been in their current roles for between one and five years, ensuring insights from both long-tenured employees and relatively newer staff members. While most participants (89%) reported no accessibility needs, 22% identified as culturally diverse, including individuals from non-English speaking backgrounds (NESB) and those with conditions such as ADHD. The diversity in job functions was balanced, with the most common roles being operational services (19%), followed by research/operations support (15%) and research managers (15%). This range of participants ensured a broad spectrum of perspectives on M365 Copilot's integration into research workflows and operational processes.

### Interview Data Analysis

The data analysis for this study was conducted using thematic analysis, a qualitative research method that systematically identifies, analyses, and interprets patterns within textual data (Clarke and Braun 2017). This method was chosen for its ability to provide in-depth perspectives into user experiences, allowing researchers to capture recurring themes while remaining grounded in participant perspectives. Thematic analysis was conducted in multiple stages to systematically identify patterns and key themes within participant responses.

The analysis began with the aggregation of all responses from the 27 participants across the 32 structured interview questions. Each response was transcribed, reviewed, and organised to ensure consistency and completeness. Researchers then familiarised themselves with the data through repeated readings, enabling an initial sense of emerging patterns. The next step involved an inductive coding process, where individual responses were examined line by line, generating open codes based on notable ideas, concerns, and insights. These initial codes were then grouped into broader thematic categories relevant to the study's research objectives: productivity, efficiency, effectiveness, ethical concerns, and user perception and satisfaction.

The theme of *productivity* was analysed using responses from questions 3, 4, 5, 6, 7, 9, and 10, which explored how M365 Copilot influenced task completion, workload distribution, and the ability to focus on high-value activities. These responses provided takeaways into how users perceived productivity gains in areas such as meeting summaries and email drafting while highlighting limitations in handling complex tasks.

To explore the *evolution of user expectations*, questions 1 and 2 were used to contrast initial assumptions about M365 Copilot with post-trial experiences. This analysis revealed a discrepancy between anticipated and actual performance, with participants expressing mixed levels of satisfaction depending on their specific needs and workflows.

The theme of *efficiency* was examined through questions 3, 4, 5, 6, and 7, where participants reflected on whether M365 Copilot optimised workflows, reduced cognitive load, or minimised redundant efforts. While some reported efficiency improvements in routine administrative tasks, others noted that verification and correction of AI-generated content often offset these gains.

The *effectiveness* of M365 Copilot was assessed using responses from questions 8, 9, 12, 16, and 17, which asked participants to evaluate the accuracy, reliability, and relevance of AI-generated outputs. This was particularly relevant for research-related tasks requiring critical thinking and domain-specific expertise, where the effectiveness of Copilot's suggestions was found to be inconsistent.

*Ethical concerns* emerged as a significant theme, with responses from questions 13 to 20 highlighting concerns around data privacy, AI bias, and trust in AI-driven decision-making. Participants varied in their confidence regarding M365 Copilot's handling of sensitive data, with some trusting organisational governance structures while others remained sceptical about AI transparency and control mechanisms.

Additionally, questions 21 to 31 addressed broader *ethical and governance issues*, including inclusivity, accountability, and long-term concerns about AI integration in professional environments. Responses reflected a strong call for improved oversight, clearer ethical guidelines, and structured policies to ensure responsible AI use.

To provide depth to the findings, representative supporting quotes were selected for each theme, ensuring that the analysis remained grounded in participants' perspectives. The integration of demographic perspective further enriched the analysis, illustrating how differences in job roles, experience levels, and professional contexts shaped perceptions of M365 Copilot. This structured approach ensured a robust understanding of the user experience, informing both the study's conclusions and recommendations for AI adoption in professional settings.

Figure 1 provides a summary of the research methodology, and themes used for analysis in this study. It outlines how interview data from 27 M365 Copilot trial participants was synthesised, thematically analysed, and refined into the key themes that will be discussed in details in next Results section.

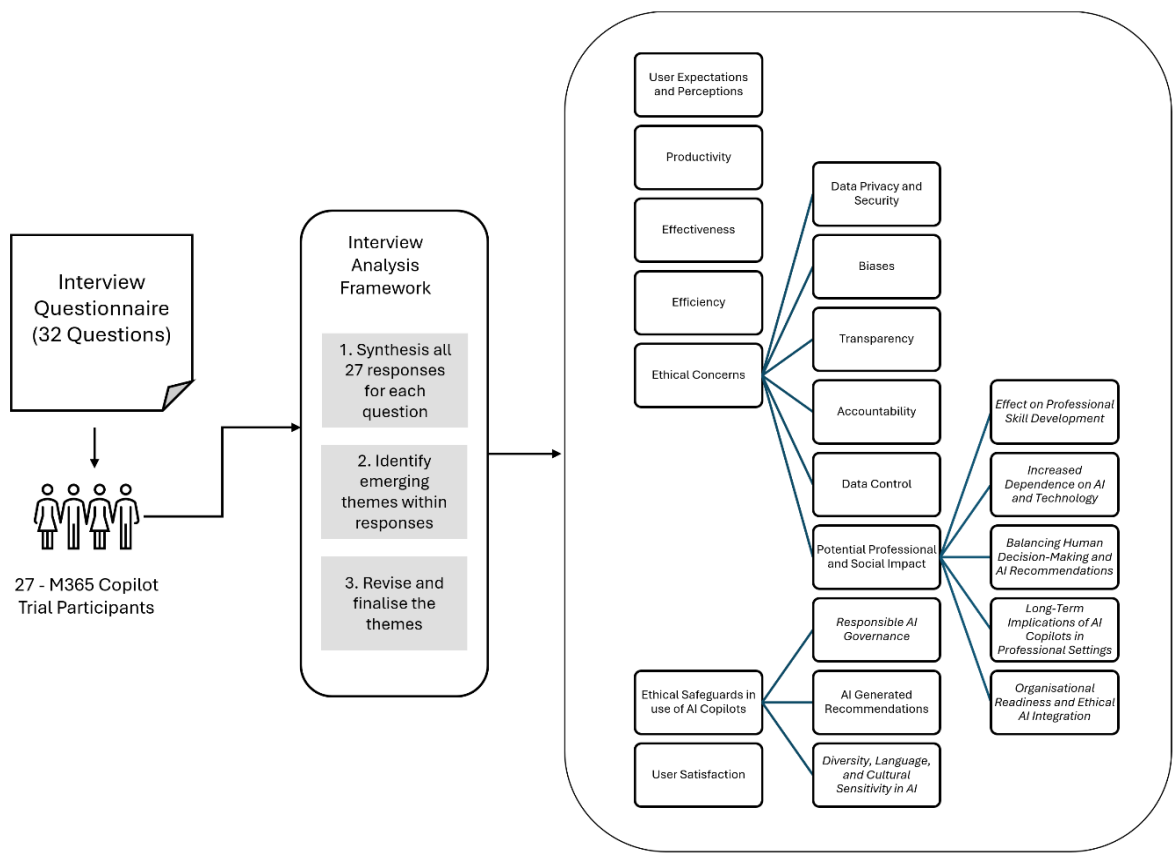

Figure 1. Research Methodology and Analysis Framework

## 4. Results

In this section we present the findings from our analysis around key themes. These themes reflect participants' experiences with Copilot during the trial and provide insight into how users evaluated its role in their workflows, decision-making processes, and professional environments.

## User Expectations and Perceptions

At the outset, participants generally held high hopes that M365 Copilot would streamline tasks, boost productivity, and possibly match ChatGPT's capabilities. They anticipated benefits across various functions—drafting emails, generating meeting summaries, and offering programming or analytical support. Some were initially sceptical, yet still curious: as one participant reflected, *"My hopes have not entirely been dashed, but it hasn't been as useful as I hoped."* Another participant, who expected it to handle a range of administrative needs, explained that *"I don't think it's as good as what I was expecting..."* showing a mixture of disappointment and surprise. Early optimism persisted for some—*"There have been definitely improvements in the usability of it"*—though they noted room for growth in ease-of-use and integration. While newer employees were keen to explore the tool's potential for immediate productivity gains, experienced staff members often had more specific aspirations, hoping for powerful, intuitive features akin to OpenAI's ChatGPT. Diverse backgrounds, including non-native English speakers and neurodiverse individuals, highlighted the importance of inclusive, culturally sensitive AI tools. There was notable excitement around time-saving features, but some participants were cautious, acknowledging possible challenges in usability and integration.

By the end of the trial, however, only 6 out of 27 participants felt their expectations had been met, with the remainder divided between those who found M365 Copilot somewhat useful and those who felt significantly let down. Several respondents praised the tool's capacity to assist with brief meetings—*"It's really good for short meetings. Overall, pretty impressed with its ability to pull out action items"*—and a handful appreciated how *"my emails are a bit better because Copilot helps in coaching my emails."* Yet concerns were prevalent about its utility beyond meetings and emails. One individual commented, *"It's kind of useless. Very good with transcription for meetings, but almost anything else...,"* while another said, *"Used it to give me summaries of meetings or webinars, saves time,"* yet found other functionalities underwhelming. Those participants who identified as Non-native English-speaking backgrounds and neurodiverse individuals stressed the importance of inclusive design and better accommodation of language barriers and cognitive differences. Whilst M365 Copilot's utility in generating meeting minutes and coaching emails was recognised, its current limitations mean that many participants remain hopeful improvements will be made to fully realise the tool's potential and deliver the seamless productivity solution initially anticipated.

## Productivity

Interviewees' perceptions of M365 Copilot's impact on their productivity varied widely. Some participants noted significant gains, with one explaining, *"I believe my productivity is increased. So I think it's been effective in that way."* Another echoed a similar sentiment, emphasising, *"It helps me a lot in terms of some meeting minutes,"* indicating that Copilot's automated note-taking and summarisation features can help users move through routine tasks more quickly.

Others observed only moderate improvements, describing how Copilot introduced additional steps but still delivered net benefits: *"There are a couple of things where it was definitely helpful, even though it added a little bit of overhead to it..."* Many of these participants noted they were adjusting to the tool's learning curve and accuracy checks.

Some respondents perceived negligible differences and stated, *"Almost no change,"* or *"I think it's the same,"* attributing their lack of improvement to the tool's incompatibility with tasks requiring nuanced thinking or specialised expertise. A few interviewees even felt that overall productivity decreased, with one remarking, *"I found it frustrating and it took longer than just doing things manually."* Another pointed out, *"Overall, it probably decreased and that's because being part of the pilot programme..."*. This illustrates how individual expectations, task complexity, and levels of tool familiarity shape whether M365 Copilot's capabilities translate into meaningful productivity gains.

## Effectiveness

The findings indicate that while M365 Copilot is perceived as valuable for routine and structured tasks, it falls short in driving significant innovation or creative solutions. Many users expressed scepticism about its capabilities in truly novel problem-solving, pointing out limitations in the quality and completeness of its outputs. One participant remarked, *"That would actually be something that I wouldn't trust it to do, knowing*

*its limitations."* At the same time, there are select instances—especially in coding efficiency, basic information retrieval, and drafting communications—where Copilot's assistance is viewed as somewhat helpful. Another interviewee reflected, *"It's a different way of finding information on the web. I think that's probably what I found a bit more innovative."* Respondents with extensive organisational experience generally see M365 Copilot as a tool for improving productivity and efficiency rather than sparking innovation, whereas newer employees may be more adaptable but sometimes lack the organisational context to leverage Copilot fully. Non-native English speakers and participants with accessibility needs note that, while Copilot can be supportive for certain routine tasks, it does not substantially advance more creative or complex endeavours. As one user put it, *"No, I wouldn't say anything innovative. I'm still the brains of the operation."*

Several respondents emphasised that M365 Copilot is most effective when used for summarisation, drafting, meeting-related activities, and technical troubleshooting. They appreciate the way it can condense documents or web content into actionable points, assist with writing initial drafts, and simplify workflows by generating meeting summaries and action items. *"I've used it extensively in relation to meetings,"* noted one participant, while another said, *"I don't know that I've given it a task that was particularly innovative, but I've used it... drafting letters..."* This functionality also extends to code interrogation, troubleshooting, and legal document analysis, saving users time in validating or refining search results. As one interviewee explained, *"Copilot... give me some code for this. And it's significantly faster than just trolling through Google..."* Interest in AI-driven solutions remains high, particularly among those who have prior experience with tools like ChatGPT, but users across the organisation voice ongoing concerns about bias and the necessity for robust ethical frameworks. Cultural and linguistic diversity within the user base highlights the importance of inclusive AI that can accommodate different communication styles and accessibility requirements.

Nonetheless, participants identified clear limitations in Copilot's performance on more intricate or domain-specific tasks, such as handling complex datasets in Excel, producing complete final drafts for PowerPoint and Word, and generating emails with the right balance of formality and conciseness. While one user noted, *"It did help me manipulate Excel in a faster way... so it automatically put something into a table..."* others encountered more specialised requirements and concluded, *"Copilot can't do it."* In these cases, Copilot's suggestions were often seen as rigid, overly verbose, or misaligned with the specialised demands of the user's role. Employees with longer tenures in research or business development roles tend to rely on established workflows that do not always align well with Copilot's current capabilities, while newer staff may be open to experimentation yet still find the tool's learning curve steep in certain areas. Accessibility considerations—including language barriers and conditions like ADHD—further highlight the need for Copilot to evolve in user-friendliness and adaptability. Ultimately, M365 Copilot's strengths lie in expediting familiar processes, whereas complex, innovative problem-solving and advanced data manipulation remain areas requiring more human oversight and expertise.

### Efficiency

Users' perceptions of efficiency reflect a broad spectrum of experiences. Several interviewees praised Copilot's ability to streamline meeting-related tasks, with one remarking, *"It has saved me a bit of time. Particularly... I usually take notes during meetings,"* while another highlighted the quick turnaround: *"It's really good because it will summarise and give you the action items really quickly."* These views emphasise Copilot's strengths in automating repetitive tasks and distilling large amounts of information into more manageable components.

A few respondents expressed disappointment, commenting, *"I was expecting more,"* and reverting to other AI (such as ChatGPT) or manual processes. Others emphasised, *"Hasn't changed,"* or *"Very little change,"* indicating that Copilot had not become integral enough to transform their daily routines. For these participants, limited functionality or persistent integration issues meant efficiency remained roughly the same.

These reflections highlight that M365 Copilot's impact on productivity and efficiency is very much rooted in user perception. Where Copilot's features align with an individual's job requirements—particularly around summarisation, note-taking, and repetitive document tasks—its benefits can be considerable. Yet for more complex, creative, or highly specialised workloads, its influence can be limited or even counterproductive. Ultimately, how each user engages with Copilot and in what context largely determines whether they perceive improvements in their work processes.

### Ethical Concerns

Respondents' feedback on Copilot reveals a range of ethical considerations that can be broadly categorised into data privacy and security, transparency, biases, accountability, data control, and potential professional and social impact. Below we discuss each of these ethical concerns:

#### Data Privacy and Security

Participants hold varied perspectives on data privacy and security measures, often balancing trust in Organisation's safeguards with a degree of caution about potential vulnerabilities. Several users expressed confidence in the system, pointing out that it is *"all stored internally"* and emphasising that *"we're using our own instance of CoPilot. There's nothing that's leaving the organisation."* Another participant adds, *"I felt totally safe... there was always that concern of sharing confidential information [with public AI], because you don't know where it's going to end up,"* but they now feel more reassured by Copilot's internal setup. Likewise, others take a similarly trusting stance—*"I am quite trusting. I assume if [Organisation] has approved this trial, they've done all the checks and balances that privacy is appropriate,"*—reiterating the belief that rigorous internal protocols are in place.

Nonetheless, some respondents harbour reservations, suggesting that the existing measures may not fully address every privacy concern. As one explains, *"I think they're good. Start... they might need some extra work... mostly to do with the Privacy parts,"* while another shares, *"I think they're a good start. I have some concerns... I never thought I had access to [this data], so... what's the potential?"* Others highlight the necessity for transparent guidelines regarding who can access what information, cautioning that they *"take extra precautions... [not to] accidentally leak any information."* Another user mentions adding a disclaimer in their emails about Copilot use and discovering, *"it was able to access my OneDrive folder... So I'm not really sure, but I was reassured... yet also a bit concerned."* These viewpoints highlight a tension between recognising Copilot's benefits and questioning whether all security protocols and user controls are sufficiently robust.

Many participants accept the organisation's assurances—*"I look, I've read through all the documents... I'm comfortable with it... I feel quite confident in security,"*—while acknowledging that incidents of inadvertent access or data sharing would undermine trust if they occurred. Several respondents also reflect on how limited or role-specific knowledge graphs can be both *"good and bad,"* sometimes restricting necessary collaboration or, conversely, risking unauthorised visibility. Calls for more detailed instructions and training are common, with one noting, *"We need to have some guidelines around what we do with these recordings,"* especially in contexts like meeting transcription or note-taking. Ultimately, while there is an undercurrent of trust in CSIRO's capacity to protect sensitive information, these varied experiences highlight the importance of ongoing communication, user education, and continuous refinements to data security measures.

#### Biases

Most respondents did not observe any overt bias in M365 Copilot's functionality or outputs. Many users explained that their interactions with Copilot were largely task-driven, involving summarisation, rewording, or data analysis rather than open-ended content generation, which may have limited their exposure to potential biases. One respondent pointed out, *"I haven't given it a task where it could do that... mine was always just analysing data and stuff,"* while another noted, *"I'm using it more for busy activities, and all the summaries I've received have not been elaborated."* This suggests that for structured tasks involving factual information, Copilot appears relatively neutral in its responses.

However, some subtle biases did emerge. One participant noted that Copilot exhibited *"a very clear bias towards particular publishers of journals, so lots and lots of [publisher name] papers... not always what I'd first go to,"* while adding that it *"limited [cultural] context"* in favour of international sources, such as reports from the International Renewable Energy Agency. This raises concerns about whether Copilot prioritises information from certain databases over others, which may impact users who require regionally relevant findings. Another user reflected on whether Copilot's *"focus on the topic and not to steer outside"* might restrict exposure to alternative viewpoints, making its responses feel overly narrow.

Language preferences were another area of concern. Several participants pointed out that Copilot defaults to American English spelling, with one stating, *"I think it defaults to American English... that's a bit of a pain,"* and another adding, *"Even when I set it to [country] English, it still doesn't seem to recognise it."* While this may seem minor, it reflects broader limitations in localisation, which could impact inclusivity and user experience. Another respondent also mentioned *"generalised responses"* when discussing responsible AI, noting that when they prompted Copilot about women's roles in AI ethics think tanks, it provided stereotypical explanations such as *"women are more caring and empathetic."* This suggests that while Copilot may not overtly promote biases, it can sometimes reflect generalised societal narratives that require refinement.

While the majority of users did not encounter explicit bias, a few instances highlight areas where improvements could be made. Ensuring Copilot sources a diverse range of information, improving localisation settings, and refining how it presents certain topics will be important steps in mitigating subtle biases and ensuring a more balanced and inclusive AI tool.

### Transparency

Respondents provided mixed feedback regarding M365 Copilot's transparency, with many appreciating its ability to reference sources while others found its explanations lacking in detail, particularly for certain tasks. Several users highlighted that Copilot provides clear citations, helping them verify information. One participant stated, *"Given the example of, say, looking at a summary from a meeting, that's been very transparent because you get little references,"* while another added, *"It always makes a reference to where the summary comes from, which is very good transparency."* Similarly, participants using Copilot for web searches or research tasks found its approach to referencing useful, with one noting, *"It gives a bit of a detailed description of where it pulled data from to generate sentences and how it formed its paragraphs."* These features were generally well received, particularly for meeting summaries and content retrieval tasks, reinforcing the tool's reliability when sourcing external information.

However, some respondents pointed out gaps in Copilot's transparency, particularly in content generation and proofreading. One user commented, *"I don't know that I actually saw it explain any of its suggestions or whether it would just do what I asked it to,"* while another mentioned, *"I don't recall seeing explanations of suggestions, but I'm not sure if that's different if you're using it for research, data processing type stuff."* Others noted that Copilot sometimes *"sort of gives you half an answer and then goes off into irrelevance,"* or *"quite often it'll come back with 'sorry, I can't do that,' but it doesn't say why."* These limitations made it difficult for users to fully understand the basis of Copilot's recommendations, leading to a need for additional clarification and validation. Some users also reported inconsistency in the quality of references, with one explaining, *"It gives the references when it does though, so that's quite good. Some things are a bit random."*

Another common concern was Copilot's tendency to lean towards positive responses, potentially limiting a balanced perspective. One respondent remarked, *"It's literally the algorithms are biased towards positive responses,"* suggesting that Copilot may not always present critical or alternative viewpoints. While some users found its explanations sufficient, one participant noted that *"it doesn't seem to be very verbose in saying why it did what it did, but you can always ask for more."* Others viewed Copilot's transparency as comparable to other AI tools, with one stating, *"Transparent, I think. I don't know. It does as good a job as any, any kind of chatbot type thing."* Meanwhile, those with lower expectations of AI capabilities were less critical, as one participant commented, *"I think it's not great, but I don't expect it to be good."*

While many users found Copilot's transparency adequate, particularly in meeting summaries and research applications, there is room for improvement in explaining its reasoning for content generation, proofreading, and more complex prompts. Enhancing the clarity of its explanations, addressing inconsistencies in references, and offering a more balanced presentation of information could strengthen user trust and usability.

### Accountability

User responses overwhelmingly highlight that the responsibility for ensuring the accuracy of M365 Copilot's outputs ultimately falls on the individual using the tool. Many participants stressed the necessity of verifying content before use, with one stating, *"The user. I mean, that's not saying it's a perfect solution...*

*The person who's presenting that, yeah.*" Others echoed this sentiment, emphasising that users must critically assess and refine Copilot-generated content: *"You've got to make sure it's up to you, it's your responsibility, not the AI, to make sure the report is correct."* Another participant reinforced this by explaining, *"I should be accountable for not checking the result."* This consensus suggests that while Copilot can enhance efficiency, it does not eliminate the need for human oversight, particularly in professional and research-based contexts.

Despite the strong emphasis on user responsibility, some respondents highlighted the role of organisational governance in mitigating errors. One participant described it as *"down to the governance model that's put in place supporting it,"* explaining that while users bear responsibility, *"if governance controls are not strong enough, then... people make mistakes. It's as simple as that."* Similarly, another respondent acknowledged the need for structured approval processes, stating, *"We have checks and balances in place to make sure information is accurate... When things fall through the gap, it's a matter of the person putting out the content making sure those checks are made before it goes out."* This perspective suggests a shared accountability model, where organisations must establish clear guidelines while still expecting users to exercise due diligence in verifying AI-generated content.

The prevailing sentiment among respondents is that accountability for errors rests primarily with the user, but there is also a need for strong governance frameworks and structured approval processes to minimise risks. While Copilot is seen as a valuable support tool, its limitations in recognising and correcting errors reinforce the necessity of human oversight. Ensuring clear organisational policies, improving Copilot's ability to self-correct, and promoting user awareness of its potential pitfalls will be essential in enhancing its reliability and trustworthiness in workplace settings.

### Data Control

User responses on data control revealed significant variation in how well-informed participants felt about Copilot's data usage and how much control they perceived they had over it. Some participants felt adequately informed and trusted the existing systems, with one stating, *"I believed the right controls had been put in place to manage data access."* Others found the integration with Microsoft's tools reassuring, explaining, *"Everything was integrated into my Microsoft knowledge graph, which made it straightforward and manageable."* However, many participants expressed uncertainty or a lack of clear understanding about what data Copilot could access and how to manage it effectively. One respondent noted, *"I initially thought of a bigger picture but then realised and confirmed that, yes, I had control, albeit with some initial confusion about Microsoft's involvement."* This reflects a broader theme where some users felt a sense of control primarily due to institutional trust rather than a deep personal understanding of data governance.

Few users reported feeling they had little to no control over what data Copilot accessed, citing concerns over visibility, documentation, and unexpected system behaviour. Several respondents admitted they had *"no clear view or understanding of how to control or lock down data access,"* while others assumed that Copilot had broad access to information without specific restrictions. One participant remarked, *"I just assumed it had access to everything, which is a bit concerning,"* while another expressed frustration over the lack of clarity, stating, *"There needs to be better documentation explaining what it can and can't see."* Some users were surprised by Copilot's ability to access certain files, with one recalling, *"I was caught off guard when it accessed something I wasn't expecting—it made me question how much control I actually have."* These concerns highlight the need for clearer guidance on how users can manage and limit Copilot's access to sensitive information.

Other respondents expressed ambivalence, acknowledging some level of control but remaining uncertain about its effectiveness. One participant shared, *"I assume I have control based on the initial setup, but I haven't really tested it,"* while another admitted, *"I haven't actively tried to control what it accesses, so I don't know how much control I really have."* Some users took a cautious approach, preferring to limit what they shared with Copilot, explaining, *"I take small steps in terms of what I let it see—I'd rather be safe than sorry."* Others described inconsistent experiences, with one stating, *"Sometimes it retrieves exactly what I need, other times it feels like it's seeing things it shouldn't."* These mixed responses indicate that while some control mechanisms may exist, they are not always well understood or effectively utilised by users.

The findings suggest a strong need for enhanced transparency, user education, and clearer mechanisms for managing data access. While some users trust existing privacy safeguards, a lack of clarity leaves many feeling uncertain or concerned about Copilot's level of access. Improving documentation, providing clearer opt-in and opt-out settings, and offering user-friendly controls would help bridge this gap, ensuring that all users feel confident in managing their data effectively.

## Potential Professional and Social Impact

Respondents expressed *cautious optimism* about Copilot's effect on employment, with many agreeing that while it can automate some tasks, it is unlikely to replace human roles entirely. Several participants saw Copilot as a tool that *"frees up people's time doing boring, repetitive stuff,"* allowing employees to focus on more valuable work. Others highlighted the potential for *enhancement and augmentation rather than replacement,* noting, *"There is room for enhancement or augmentation, but it won't replace everything."* However, some participants remained wary of automation's broader implications, stating, *"There is potential, which is scary and not at the same time."*

Scepticism was more pronounced among those in research and complex analytical roles, with one participant explaining, *"There's a lot of contextual stuff that I can't see it replicating."* Others acknowledged Copilot's efficiency benefits but emphasised that *automation does not mean elimination,* explaining, *"It may not replace whole roles, but parts of roles can be automated."*

From a demographic perspective, *more experienced professionals and those in managerial or research roles offered nuanced perspectives*, recognising that Copilot might shift job responsibilities rather than remove them. Those in operational and support roles were generally more sceptical, believing that *human oversight remains essential* in their workflows. Meanwhile, participants with diverse backgrounds, including non-native English speakers and those with accessibility needs, were more likely to *view Copilot positively* for its potential to improve productivity and efficiency.

*Effect on Professional Skill Development:*

The impact of Copilot on skill development varied among respondents. While a small group reported positive effects on efficiency and communication, the majority saw little to no impact, often comparing it unfavourably to more advanced AI tools. One user stated, *"Freeing up time from writing painstaking, detailed meeting notes has been helpful,"* while another noted that *"It mainly helps in how you communicate—especially in coaching emails."* However, several participants felt that Copilot had not changed their professional approach, with one remarking, *"I do not feel that Copilot has affected the development or application of my professional skills."* Others were more open to its potential, explaining, *"I was thinking about how it might help create texts like documents, but not yet."*

Early adopters and those in operational or administrative roles were *more likely to report positive changes*, particularly in *efficiency and communication*. In contrast, *research managers and business development professionals with longer tenure* were *less likely to see significant benefits,* possibly due to *established workflows* that already meet their needs. Some participants with diverse backgrounds, such as *non-native English speakers or individuals with ADHD,* noted that Copilot could *help bridge communication gaps,* suggesting that *targeted training could enhance its impact across different user groups.*

*Increased Dependence on AI and Technology:*

Most respondents stated that using Copilot had *not significantly increased their dependence on AI*, as they were already highly reliant on technology in their roles. One participant stated, *"I'm totally dependent on technology already, so probably no difference,"* while another explained, *"I don't think it necessarily introduces any new dependencies—it feels quite additive at this stage."*

However, a few users admitted to increased reliance in specific areas, particularly in *coding and troubleshooting*. One respondent explained, *"Yes, I think there's some, yeah... coding—I know I'll get there so much faster with Copilot."* Another participant noted, *"Sometimes I think in a meeting, 'Oh, Copilot will get it,' and then realise I should have written it down."* This suggests that while Copilot may not *fundamentally change overall dependency on AI*, it does create *new habits of reliance* for certain automated tasks.

Across demographic groups, perceptions of AI dependency *did not vary significantly,* reinforcing that *Copilot's impact depends more on individual workflows than broad demographic factors.*

*Balancing Human Decision-Making and AI Recommendations:*

There was a strong consensus that *human oversight remains crucial* when integrating AI into decision-making. Many participants acknowledged that AI could *generate ideas, summarise information, and streamline tasks* but *stressed that final decisions must remain with humans.* One respondent explained, *"AI can be a really powerful tool in giving you guidance… but ultimately, decisions come down to human judgment."* Another reinforced this by stating, *"AI can provide recommendations, but it should not replace human decision-making… it's important to maintain a level of human oversight."*

Concerns over *trust and reliability* were prevalent, with some users describing AI-generated content as helpful but requiring *rigorous verification.* One participant noted, *"I still think you need that human check and balance—I wouldn't be prepared to send out anything without double-checking it first."* Others shared similar sentiments, stating, *"I am always critical of the information that comes out… But the steps before that are accelerated."* Some respondents also pointed out that *regulations and clear policies are necessary* to ensure AI is used responsibly, explaining, *"There needs to be regulation and just enough information for what people use."*

*Long-Term Implications of AI Copilots in Professional Settings:*

Long-term ethical concerns raised by participants focused on privacy, human oversight, misinformation risks, and AI's potential to alter professional workflows. Several participants highlighted the risks of AI being used to make decisions without human oversight, with one stating, *"The biggest concern is over-reliance on AI, where decisions are made without human intervention."*

Others were concerned about the erosion of human communication and authenticity in professional environments. One respondent noted, *"Using AI to draft messages can make the workplace feel less human, as you realise you're not interacting with a person."*

Misinformation was another prominent concern, with one participant warning, *"AI's ability to spread biased or inaccurate information quickly is a major risk, and we need rigorous safeguards."* There was also a recognition that AI learning patterns could reinforce user biases over time, leading to a narrowed perspective of information: *"AI learns your behaviour and preferences, which could create a narrow ecosystem where we only see what we want to hear."*

*Organisational Readiness and Ethical AI Integration:*

Participants strongly recommended that organisations develop structured policies to regulate AI use, including task-specific guidelines, clear governance frameworks, and robust training programs. One participant stated, *"Firm policies should define what AI can and cannot do, ensuring human oversight in critical tasks."* Another noted the importance of communication and transparency, saying, *"Having open discussions with staff is essential to address concerns and ensure AI is suitable for specific work tasks."*

Respondents also called for stronger governance structures to ensure AI integration aligns with ethical principles. One participant emphasised, *"Rather than banning AI, organisations should co-design AI systems with appropriate risk assessments and stopgaps where human decision-making is required."* Another participant added, *"Training employees to work with AI and understand its limitations is key to its responsible implementation."*

*Addressing Emerging Ethical Issues in AI:*

While many participants did not identify specific ethical issues with M365 Copilot, others pointed out potential risks, such as AI misinterpreting data, job displacement concerns, and lack of recompense for human knowledge contributions. One respondent questioned, *"What's the recompense for people whose knowledge AI is learning from, without giving them anything back?"* Another mentioned concerns about AI systems communicating without adding real value, saying, *"We could get into a situation where Copilot is talking to another Copilot, wasting computing power with no meaningful outcome."*

## Ethical Safeguards in use of AI Copilots

This section explores the ethical dimensions surrounding the deployment of AI copilots, focusing on participants' perspectives and concerns. Key themes include responsible AI governance practices, the perceived reliability and transparency of AI-generated recommendations, and the importance of diversity, language, and cultural sensitivity in AI outputs.

*Responsible AI Governance*

Participants overwhelmingly supported the need for ethical safeguards and governance frameworks to regulate AI technologies like Copilot. Key concerns included bias in AI models, confidentiality, intellectual property rights, and the responsible use of generative AI. One respondent highlighted the bias issue in AI training data, stating, *"Bias is a huge concern, especially when looking at existing training data that embeds human bias. People presume AI is neutral, but it reflects the biases of its sources."* Another participant highlightd the need for organisational strategies, stating, *"We need an automation strategy that is ethically centred around augmenting the human experience, not replacing it."*

In high-risk fields such as research and legal compliance, confidentiality and appropriate data controls were highlighted as necessary safeguards. One participant expressed concern, saying, *"In my field, we deal with confidential information... It could be very wrongly used without the proper controls."* Some participants also pointed to intellectual property concerns, where AI systems use external data without appropriate permissions. *"Learning from artists who are not aware that their work has been used in training datasets is a huge issue,"* noted one respondent.

Participants suggested that ethical guidelines should be embedded into an organisation's AI strategy, ensuring AI complements rather than replaces human expertise. Some also advocated for clear user training to help employees understand how AI operates and its associated risks: *"There needs to be training so that users understand the way AI operates and the risks involved."*

*AI Generated Recommendations*

Most participants reported no ethical discomfort with Copilot's recommendations, primarily because their interactions were limited to non-controversial tasks like drafting content and summarising meetings. One respondent noted, *"It hasn't really given me recommendations per se, more summaries."* Another stated, *"I have not felt uncomfortable with Copilot's recommendations due to ethical concerns."* However, some respondents remained cautious, citing the potential for algorithmic bias and trust issues in decision-making. *"Human oversight in decision-making is essential to ensure AI is not operating as a 'black box' without accountability,"* one participant commented.

A small number of users expressed uncertainty about the extent of Copilot's influence, particularly in decision-making contexts. One participant remarked, *"I have seen some concerns from others, but I personally haven't encountered ethical risks that I can comment on."* This suggests that while Copilot's ethical risks may not yet be fully visible, ongoing scrutiny is necessary.

*Diversity, Language, and Cultural Sensitivity in AI*

The ability of Copilot to handle diverse languages and cultural contexts received mixed feedback. Some participants, particularly non-native English speakers, found value in Copilot's language assistance, with one respondent stating, *"This is one of the main reasons I was excited about Copilot – to use it as coaching for my emails, as English is not my first language."* Others, however, pointed out M365 Copilot's struggles with accents, specific dialects, and cultural nuances. One user shared, *"I tried to use the speech-to-text function with Copilot, and it wasn't able to pick up my instructions due to my accent."*

Several respondents were uncertain, stating they had not tested Copilot's language diversity features extensively. One participant summed up the general sentiment: *"I haven't seen a lot of evidence either way."* The findings indicate a need for further evaluation and improvement in AI's ability to support diverse linguistic and cultural needs.

## User Satisfaction with M365 Copilot

Findings from the interviews reveal that user satisfaction with M365 Copilot is mixed, reflecting both optimism about its potential and frustrations stemming from unmet expectations and current limitations.

Some participants reported high levels of satisfaction, with one noting, *"I'd say satisfied to very satisfied... you can speak to it in plain language"* and another adding, *"I'm satisfied in terms of how it helps me during my day-to-day work. Overall, I'm pretty happy with it."* Others, however, described being *"not satisfied"* or only *"slightly above neutral"* due to a lack of trust in Copilot's accuracy and scope—*"I'd say not satisfied in this in my case... there's an element of not really trusting the outputs, particularly when my role is quite detailed oriented."* These varied perspectives highlight the gap between users' hopes for robust, time-saving features and their experiences with incomplete functionalities or confusing prompts. Despite the discrepancies, many participants see promise in Copilot's future developments, with one remarking, *"There are some things I didn't expect that I'm enjoying... and some other things I had high hopes for... but it didn't quite pan out."*

When viewed through demographic lenses, similar nuances emerge. Employees with longer tenures in more strategic or managerial roles acknowledge Copilot's *"strategic potential"* but remain guarded about its current shortcomings, echoing feedback such as, *"If it evolves a bit more... it could be a bit more helpful."* Newer or more junior staff often appreciate the immediate, practical benefits—*"There were some things I didn't expect that I'm enjoying"*—yet can become frustrated by usability hurdles and limited capabilities. Cultural and linguistic diversity also plays a role, with non-native speakers stressing the importance of inclusivity in language handling, and participants with specific accessibility requirements noting the need for Copilot to be *"very specific and accurate enough"* for detail-oriented tasks. Meanwhile, a number of early adopters and tech enthusiasts are vocal about its advantages—particularly *"significant efficiencies with programming"*—but they also criticise constraints on prompts compared to more advanced AI models.

The broader user satisfaction hinges on how well Copilot aligns with everyday tasks and its potential for growth. While features like automated summarisation and drafting *"help me during my day-to-day work"* or *"save time in research"* garner praise, issues with inaccurate outputs, data governance, and underdeveloped features temper broader enthusiasm. Many users fall somewhere between *"neutral"* and *"somewhat satisfied,"* underscoring the importance of refining Copilot's functionality and inclusivity. By addressing these concerns—improving accuracy, offering more tailored training, and expanding capabilities—M365 Copilot could see higher satisfaction and more widespread adoption.

## 5. Discussion

In this section, we discuss implications of our findings in broader context.

### Rise of AI Copilots

The proliferation of AI copilots marks a significant evolution in how organisations integrate AI into professional environments. AI copilots are designed to augment human intelligence by providing real-time assistance in drafting, summarisation, coding, and data retrieval. The professional landscape is now increasingly reliant on AI-assisted tools across various domains.

Despite their rapid adoption, AI copilots are not without limitations. Findings from our interview study suggest a divergence between expectations and actual capabilities of AI Copilots. As AI copilots continue to evolve, their role in professional settings will likely expand, but their current impact remains incremental rather than transformative. While they excel at optimising repetitive and structured tasks, their effectiveness in higher-order cognitive tasks remains inconsistent. Organisations must therefore approach their adoption with a critical perspective, recognising both their potential and their constraints. The growing integration of AI copilots into enterprise workflows signals a shift toward AI-augmented decision-making, but the technology's ability to fully integrate into complex professional environments without significant human oversight remains an open question.

### Productivity Paradox

One of the key findings of the study was that while Copilot helped automate routine administrative tasks, its effectiveness in higher-order cognitive work remained inconsistent. Users frequently found themselves needing to double-check and manually refine Copilot's suggestions, undermining the time savings AI was supposed to provide. This aligns with the *productivity paradox* (Brynjolfsson 1993)*,* where significant investments in advanced technology do not always lead to the expected productivity gains. The study revealed that Copilot's impact was often offset by usability challenges, lack of domain-specific knowledge, and the necessity for human oversight. These factors resulted in a shift of effort rather than a clear reduction in workload, reinforcing the notion that AI copilots currently act as productivity enhancers rather than productivity transformers.

The *Paradox of Workplace Productivity* (Schwartz 2010; Fuller 2016), which occurs when individual efficiency improvements do not necessarily translate into organisational gains, provides further insight into the mixed results of M365 Copilot's adoption. While many users in the study reported that Copilot helped with individual tasks such as summarising meetings, drafting emails, and formatting documents, these benefits did not consistently scale to broader organisational productivity improvements. This discrepancy highlights a fundamental challenge: AI copilots can streamline specific workflows, but their success in enhancing overall efficiency depends on how well they are integrated into an organisation's broader systems, policies, and work culture.

### Trust Dynamics

The interviews revealed a nuanced trust dynamic in how participants perceived ethical concerns related to M365 Copilot. A recurring theme was that participants expressed greater confidence in *CSIRO* and its internal IT governance teams than in Microsoft when it came to handling data privacy, security, and ethical oversight. This institutional trust played a significant role in shaping user perceptions, with many participants feeling reassured that AI adoption was being managed within a familiar and accountable framework. As a result, ethical concerns, such as biases in AI outputs, data governance, and the responsible use of AI-generated content, were less frequently raised as primary adoption barriers than might have been expected in an external deployment scenario.

This finding suggests that while AI tools like M365 Copilot may be accepted at a functional level, their success in an organisation depends significantly on the credibility and transparency of the implementing body. Participants indicated that their trust in CSIRO's governance standards and data protection measures influenced their comfort level in using Copilot, even when uncertainties about AI decision-making and ethical implications persisted. This aligns with broader research on AI adoption (Lukyanenko, Maass, and Storey 2022), which highlights that trust in the deploying institution is as critical as trust in the technology itself.

The findings also highlight that users are not simply evaluating AI on technical performance alone, but also on whether the organisation's values, safeguards, and ethical commitments align with their own expectations. This suggests that for organisations considering AI copilots, establishing clear, institution-led AI governance frameworks—including transparency in decision-making, user training on ethical AI use, and robust accountability structures—will be crucial to ensuring long-term user acceptance and trust. Without these foundational elements, even a well-functioning AI system may struggle with user engagement due to perceived ethical and security risks.

### User Expectations vs. Actual Capabilities of M365 Copilot

A recurring theme in user feedback was the gap between expectations and the actual capabilities of Microsoft 365 Copilot. Many users anticipated that the AI would function as an intuitive and near-autonomous assistant, but their experience revealed significant limitations. Participants found that Copilot was not well suited for complex problem-solving, and some expressed concerns about bias in the sources it prioritised, particularly in research-related outputs. Others noted that Copilot lacked explanatory depth, often providing summaries without clear justifications for its suggestions, making it difficult to trust its recommendations. These discrepancies between user expectations and AI performance suggest that better communication and training around the tool's capabilities are essential. Managing trust and expectations remains a significant challenge in AI adoption, as users must balance their reliance on AI with an understanding of its inherent limitations.

### Known-Unknowns of AI Copilots

The trial of M365 Copilot within CSIRO presents a complex interplay of knowns and unknowns, reflecting both tangible benefits and unresolved uncertainties in AI implementation. The *known-knowns* include clear productivity gains in specific areas such as meeting summarisation, email drafting, and basic data retrieval, where AI automation has streamlined workflows and reduced manual effort. Users generally acknowledged that Copilot accelerates routine tasks and facilitates information access, making it a useful addition to their daily operations. However, these benefits are incremental rather than transformative, as Copilot's effectiveness is often limited to structured and well-defined tasks rather than complex problem-solving or innovative workflows.

Despite these clear advantages, several *known-unknowns* emerged in user experiences, highlighting challenges that are recognised but not yet fully understood. Data security and governance were among the most prominent concerns. While some participants trusted CSIRO's internal safeguards, others expressed uncertainty over how much control they had over what data Copilot could access. The need for greater transparency and clearer governance structures was repeatedly emphasised, as users sought reassurances on privacy, compliance, and information security. Similarly, bias and accuracy concerns were noted by some participants, who observed Copilot's tendency to favour specific journal publishers, default to generalised responses, or struggle with nuanced prompts. These findings indicate that while AI copilots can be trusted for routine automation, their interpretation of content and contextual judgment still require human oversight to ensure accuracy and fairness.

Another significant unknown is how AI copilots will influence professional skill development and long-term workforce dynamics. Some participants highlighted that Copilot helped improve efficiency in drafting and summarising content, but others raised concerns that excessive reliance on AI could erode critical thinking, reduce information retention, and create an over-dependence on automation. These fears align with broader discussions on AI in the workplace, where the risk of deskilling due to AI-generated content needs to be weighed against the benefits of enhanced efficiency and knowledge retrieval. Moreover, the impact of AI copilots on collaboration, decision-making processes, and workplace culture remains an evolving question, requiring ongoing evaluation and adaptation.

Beyond the recognised uncertainties, there are also *unknown-knowns*, aspects of Copilot's capabilities that may exist but are not yet fully utilised or understood by users. Some participants acknowledged advanced functionalities they had not explored, while others noted Copilot's potential for creative applications but lacked confidence in its ability to handle non-routine tasks. This suggests a gap between what the AI tool can offer and what users currently leverage, emphasising the importance of targeted training and greater user awareness. Additionally, potential biases in AI outputs may go unnoticed, particularly in content summarisation and recommendations, which could subtly shape decision-making over time.

Finally, the *unknown-unknowns* represent the long-term, unpredictable effects of AI copilots in professional settings. These include future ethical dilemmas, shifts in workplace hierarchies, and unforeseen dependencies on AI for knowledge work. Concerns about the evolving role of AI in decision-making, the potential homogenisation of ideas, and hidden vulnerabilities in AI-generated content remain speculative but warrant continued scrutiny. The introduction of Copilot is not merely a technological upgrade but a structural change in how work is performed, evaluated, and managed. Its full impact will only become apparent over time as workflows, expectations, and ethical considerations evolve in response to increasing AI integration.

To address these uncertainties, organisations must adopt a proactive approach to AI governance, ethical oversight, and continuous user engagement. Establishing clear policies on data handling, providing comprehensive user training, and promoting a culture of AI literacy will be critical in ensuring that AI copilots augment rather than undermine human expertise. By identifying and mitigating risks early, organisations can leverage AI for meaningful productivity gains while safeguarding ethical, professional, and organisational integrity.

### Changing Landscape of AI

The global landscape of AI adoption is evolving at an unprecedented pace, driven by rapid advancements in artificial general intelligence (AGI) (Goertzel 2014) and AI agents (Kolt 2025) that extend beyond traditional copilots. The competition in this space is intensifying, with emerging international AI players such as DeepSeek-V3[7] challenging established AI ecosystems like OpenAI's ChatGPT or M365 Copilot. Unlike existing AI copilots, which primarily function as assistive tools, future AI systems—including AGI-driven agents—aim to operate with greater autonomy, contextual reasoning, and self-improving capabilities. These developments will fundamentally alter the way AI copilots are integrated into workflows, shifting from task-based automation toward more dynamic, decision-making roles.

As AI copilots become more sophisticated, their increasing integration into professional environments makes it harder for employees to function without them, raising concerns about AI dependence. While AI copilots already enhance efficiency in routine tasks, the next wave of AI agents will move toward independent task execution, reducing human intervention in problem-solving, research synthesis, and strategic planning. This shift is not merely about replacing manual effort but reshaping knowledge work, where AI handles lower-order cognitive functions while humans focus on judgment, creativity, and strategic oversight. However, despite these advancements, true AGI remains a long-term aspiration, and current AI agents still require human validation to mitigate risks related to bias, misinformation, and interpretability.

The international competition between AI ecosystems further accelerates this transformation. DeepSeek's advancements in AGI-oriented AI models signal an impending shift in how AI copilots and assistants will function. Unlike M365 Copilot, which is largely an augmentation tool for Microsoft's ecosystem, AI agents developed by competing AI firms are pushing toward more autonomous decision-making capabilities. The question for organisations is no longer whether to adopt AI copilots but how to strategically integrate AI agents in ways that align with governance, workforce dynamics, and ethical considerations. The rapid development of multimodal AI systems—capable of text, image, and voice-based reasoning—means that organisations must prepare for a future where AI agents operate alongside employees in a far more embedded and autonomous manner than current copilots allow.

## Organisational Return on Investment

For organisations assessing the adoption of AI copilots such as M365 Copilot, the return on investment (ROI) is a critical factor. The promised benefits of copilots include enhanced efficiency in repetitive tasks, reduced cognitive burden on employees, and long-term cost savings through workflow automation. In many cases, Copilot's ability to summarise content, automate email generation, and assist with coding has demonstrated clear value. However, these benefits must be weighed against potential risks and implementation costs.

One of the primary challenges in realising ROI is the need for continuous human oversight. While AI copilots reduce manual effort, their outputs require validation, refinement, and fact-checking, sometimes offsetting the expected efficiency gains. Short-term productivity improvements may be marginal due to the learning curve, integration issues, and trust-building processes that AI tools must undergo before they become indispensable. Moreover, the licensing and implementation costs for enterprise AI solutions, such as M365 Copilot, are substantial. This raises the question of whether the organisation's specific workflow challenges align with the capabilities of Copilot or whether alternative AI solutions may provide better value.

Another key consideration is workforce readiness. Organisations must assess whether employees are sufficiently prepared to adopt AI-assisted workflows. Even with advanced capabilities, AI copilots and agents cannot replace human expertise in nuanced decision-making, creativity, and contextual problem-solving. Companies must invest in employee training, AI literacy programs, and adaptive governance frameworks to ensure that AI functions as an augmentation tool rather than a disruptive force.

Ultimately, the decision to invest in AI copilots should be guided by a structured cost-benefit analysis. Organisations should evaluate whether AI copilots address an actual productivity bottleneck or introduce unnecessary complexity, whether the workforce and leadership are aligned in their understanding of AI capabilities and limitations, and what level of human oversight and ethical governance is required to ensure AI-generated content is reliable. They must also consider whether short-term implementation challenges outweigh long-term efficiency gains and whether an alternative AI model—such as a more autonomous AI agent—would better serve the organisation's needs.

---

[7] https://www.deepseek.com/

### Recommendations

For organisations considering the adoption of AI copilots, a structured decision-making framework is essential. While AI copilots have the potential to enhance efficiency, particularly in administrative and information-processing tasks, their limitations must be acknowledged. Potential advantages include streamlining summarisation, assisting non-native English speakers, and supporting AI-assisted problem-solving in coding and research. However, these benefits must be weighed against potential drawbacks, such as the need for continuous human oversight, the risk of biased or inaccurate outputs, and high licensing costs. Before committing to Copilot, organisations should assess whether their workflows align with its capabilities, whether employees are adequately trained to use AI effectively, and whether AI-generated outputs meet the organisation's quality standards.

To ensure effective AI adoption, organisations should establish clear guidelines for Copilot's use. Task-specific policies should delineate where AI is appropriate, ensuring that human oversight remains a critical component in decision-making. Transparency and accountability mechanisms should be in place to address ethical concerns, such as bias, misinformation, and data security. Training programs should be implemented to equip employees with the necessary skills to work alongside AI, promoting a culture of responsible AI usage. Finally, governance structures should be designed to monitor AI performance, ensuring that its integration into daily operations aligns with organisational objectives.

The rise of AGI and AI agents means that the current generation of copilots, including M365 Copilot, will soon be eclipsed by more advanced, autonomous AI assistants. Organisations must take a forward-thinking approach, ensuring that AI investments today align with the rapidly evolving technological landscape. By making strategic, well-governed decisions on AI adoption, organisations can avoid the pitfalls of short-term efficiency hype and instead build a resilient, AI-enabled workforce that is prepared for the next wave of intelligent automation.

## 6. Threats to Validity

### Internal Validity

The voluntary nature of participation in the study may have resulted in self-selection bias, where individuals who were more engaged with AI technologies or had particularly strong opinions, either positive or negative, were more likely to participate. This could have led to a non-representative sample, potentially skewing feedback toward more extreme perspectives. Additionally, participants might have altered their responses based on perceived expectations or social desirability, especially in a one-on-one interview setting, leading to potential bias in self-reported experiences. Efforts were made to minimise this by assuring participants of anonymity and confidentiality to encourage honest and unbiased responses.

### External Validity

With only 27 interview participants, the findings may not fully represent the broader population of users within CSIRO. While the persona-based participant selection aimed to capture a diverse range of experiences across different roles, the relatively small sample size limits the generalisability of results to other organisations or industries. Furthermore, CSIRO's unique work environment, institutional policies, and existing technology infrastructure may have influenced users' interactions with M365 Copilot, making it difficult to generalise findings to different organisational settings where AI adoption dynamics may differ.

### Construct Validity

The study relied on structured interviews with predefined questions, which, while designed to capture user perceptions comprehensively, may not have fully accounted for the complexity of participants' experiences with M365 Copilot. The qualitative nature of the data introduces the potential for subjective interpretation, both from participants when expressing their views and from researchers when analysing responses. Additionally, while the questions aimed to explore productivity, efficiency, effectiveness, ethical concerns, and user satisfaction, participants' understanding of these concepts may have varied, potentially affecting consistency in responses.

### Addressing the threats to validity

To mitigate these validity concerns, a persona-based approach was employed during participant selection, ensuring representation across different roles, levels of experience, and work contexts. The interview questions were designed to be open-ended and neutral to minimise response bias and allow participants to express genuine experiences. Multiple researchers were involved in the thematic analysis to reduce individual interpretation bias and enhance the reliability of findings. Additionally, participants were assured of confidentiality and the voluntary nature of their involvement, reducing the likelihood of socially desirable responses. By implementing these strategies, the study aimed to strengthen the internal, external, and construct validity, providing a more balanced and reliable understanding of M365 Copilot's impact in a research-focused organisational setting.

## 7. Conclusion

The interview study provided nuanced qualitative results into user experiences with M365 Copilot, revealing both its potential and its limitations within CSIRO, a scientific research institution. While many participants entered the trial with high expectations, their post-trial reflections presented a more complex picture of mixed satisfaction. Users acknowledged Copilot's value in streamlining meeting summaries, drafting emails, and assisting with document creation, yet they also encountered significant usability challenges, integration barriers, and inconsistencies in its performance. The findings highlighted the reality that while AI-assisted tools can enhance productivity in structured tasks, their effectiveness in complex, domain-specific work remains uneven, requiring substantial human oversight.

Ethical considerations emerged as a critical theme, with participants expressing greater trust in CSIRO's internal governance and IT department over Microsoft in handling data privacy and security concerns. This highlights the pivotal role of institutional integrity, transparency, and governance in shaping user acceptance of AI technologies. The study reinforced the necessity of robust data management frameworks, clear ethical guidelines, and ongoing training to ensure AI tools like Copilot serve as responsible augmentations rather than sources of risk or inefficiency.

These findings provide a valuable roadmap for CSIRO, the broader research community, and other organisations considering AI copilots. The study demonstrates that AI adoption must be carefully managed to balance automation with human oversight, ensuring that AI tools serve as effective augmentations rather than sources of risk or inefficiency. Future advancements in AI copilots will likely enhance their contextual reasoning and integration capabilities, but organisations must remain proactive in establishing responsible AI policies, investing in user training, and refining their deployment strategies. By taking a thoughtful and structured approach, organisations can harness the benefits of AI while mitigating risks, ensuring that copilots contribute meaningfully to productivity, efficiency, and ethical AI governance.

## References

Adetayo, Adebowale Jeremy, Mariam Oyinda Aborisade, and Basheer Abiodun Sanni. 2024. 'Microsoft Copilot and Anthropic Claude AI in education and library service', *Library Hi Tech News*.

Bano, Muneera, Rashina Hoda, Didar Zowghi, and Christoph Treude. 2024. 'Large language models for qualitative research in software engineering: exploring opportunities and challenges', *Automated Software Engineering*, 31: 8.

Bano, Muneera, Didar Zowghi, Pip Shea, and Georgina Ibarra. 2023. 'Investigating responsible AI for scientific research: an empirical study', *arXiv preprint arXiv:2312.09561*.

Bano, Muneera, Didar Zowghi, and Jon Whittle. 2023. 'AI and Human Reasoning: Qualitative Research in the Age of Large Language Models', *The AI Ethics Journal*, 3.

Bano, Muneera, Didar Zowghi, Jon Whittle, Liming Zhu, Andrew Reeson, Rob Martin, and Jen Parsons. 2024. 'Survey Insights on M365 Copilot Adoption', *arXiv preprint arXiv:2412.16162*.

Bejarano, C, AL Rodríguez Vázquez, A Colomer, J Cantero, A Ferreira, L Moens, Alexandre Duchevet, Jean-Paul Imbert, and Théo De La Hogue. 2022. 'HARVIS: dynamic rerouting assistant using deep learning techniques for Single Pilot Operations (SPO)', *Transportation research procedia*, 66: 262-69.

Bird, Christian, Denae Ford, Thomas Zimmermann, Nicole Forsgren, Eirini Kalliamvakou, Travis Lowdermilk, and Idan Gazit. 2022. 'Taking Flight with Copilot: Early insights and opportunities of AI-powered pair-programming tools', *Queue*, 20: 35-57.

Brynjolfsson, Erik. 1993. 'The productivity paradox of information technology', *Communications of the ACM*, 36: 66-77.

Chen, Wei-Yu. 2024. 'Intelligent Tutor: Leveraging ChatGPT and Microsoft Copilot Studio to Deliver a Generative AI Student Support and Feedback System within Teams', *arXiv preprint arXiv:2405.13024*.
Clarke, Victoria, and Virginia Braun. 2017. 'Thematic analysis', *The journal of positive psychology*, 12: 297-98.
Coyle, Jeff, and Stephen Jeske. 2023. 'The rise of AI copilots: How LLMs turn data into actions, advance the business intelligence industry and make data accessible company-wide', *Applied Marketing Analytics*, 9: 207-14.
Dakhel, Arghavan Moradi, Vahid Majdinasab, Amin Nikanjam, Foutse Khomh, Michel C Desmarais, and Zhen Ming Jack Jiang. 2023. 'Github copilot ai pair programmer: Asset or liability?', *Journal of Systems and Software*, 203: 111734.
Dell'Acqua, Fabrizio, Edward McFowland III, Ethan R Mollick, Hila Lifshitz-Assaf, Katherine Kellogg, Saran Rajendran, Lisa Krayer, François Candelon, and Karim R Lakhani. 2023. 'Navigating the jagged technological frontier: Field experimental evidence of the effects of AI on knowledge worker productivity and quality', *Harvard Business School Technology & Operations Mgt. Unit Working Paper*.
Fuller, Ryan. 2016. 'The paradox of workplace productivity', *Harvard Business Review*: 1-6.
Goertzel, Ben. 2014. 'Artificial general intelligence: concept, state of the art, and future prospects', *Journal of Artificial General Intelligence*, 5: 1.
Hajkowicz, Stefan, Alexandra Bratanova, Emma Schleiger, and Claire Naughtin. 2023. "Australia's artificial intelligence ecosystem: Catalysing an AI industry." In.: CSIRO, Canberra. The authors would like to thank the many experts and ....
Hannon, Brendan, Yulia Kumar, Dejaun Gayle, J Jenny Li, and Patricia Morreale. 2024. 'Robust Testing of AI Language Model Resiliency with Novel Adversarial Prompts', *Electronics*, 13: 842.
Hayawi, Kadhim, and Sakib Shahriar. 2024. 'AI Agents from Copilots to Coworkers: Historical Context, Challenges, Limitations, Implications, and Practical Guidelines'.
Horne, Dwight. 2023. "PwnPilot: Reflections on Trusting Trust in the Age of Large Language Models and AI Code Assistants." In *2023 Congress in Computer Science, Computer Engineering, & Applied Computing (CSCE)*, 2457-64. IEEE.
Kolt, Noam. 2025. 'Governing ai agents', *arXiv preprint arXiv:2501.07913*.
Kytö, Miska. 2024. 'Copilot for Microsoft 365: A Comprehensive End-user Training Plan for Organizations'.
Li, Fan, Shanshan Feng, Yuqi Yan, Ching-Hung Lee, and Yew Soon Ong. 2024. 'Virtual Co-Pilot: Multimodal Large Language Model-enabled Quick-access Procedures for Single Pilot Operations', *arXiv preprint arXiv:2403.16645*.
Lovell, CA Knox. 1993. 'Production frontiers and productive efficiency', *The measurement of productive efficiency: techniques and applications*, 3: 67.
Lukyanenko, Roman, Wolfgang Maass, and Veda C Storey. 2022. 'Trust in artificial intelligence: From a Foundational Trust Framework to emerging research opportunities', *Electronic Markets*, 32: 1993-2020.
Menz, Bradley D, Nicole M Kuderer, Stephen Bacchi, Natansh D Modi, Benjamin Chin-Yee, Tiancheng Hu, Ceara Rickard, Mark Haseloff, Agnes Vitry, and Ross A McKinnon. 2024. 'Current safeguards, risk mitigation, and transparency measures of large language models against the generation of health disinformation: repeated cross sectional analysis', *bmj*, 384.
Moroz, Ekaterina A, Vladimir O Grizkevich, and Igor M Novozhilov. 2022. "The Potential of Artificial Intelligence as a Method of Software Developer's Productivity Improvement." In *2022 Conference of Russian Young Researchers in Electrical and Electronic Engineering (ElConRus)*, 386-90. IEEE.
Nazareno, Luísa, and Daniel S Schiff. 2021. 'The impact of automation and artificial intelligence on worker well-being', *Technology in Society*, 67: 101679.
Neely, Andy, Mike Gregory, and Ken Platts. 1995. 'Performance measurement system design: A literature review and research agenda', *International journal of operations & production management*, 15: 80-116.
'Office of Science and Technology Policy (2023), The National Artificial Intelligence Research and Development Strategic Plan.'. 2023. https://www.whitehouse.gov/wp-content/uploads/2023/05/National-Artificial-Intelligence-Research-and-Development-Strategic-Plan-2023-Update.pdf.
Pearce, Hammond, Baleegh Ahmad, Benjamin Tan, Brendan Dolan-Gavitt, and Ramesh Karri. 2022. "Asleep at the keyboard? assessing the security of github copilot's code contributions." In *2022 IEEE Symposium on Security and Privacy (SP)*, 754-68. IEEE.
Peng, Sida, Eirini Kalliamvakou, Peter Cihon, and Mert Demirer. 2023. 'The impact of ai on developer productivity: Evidence from github copilot', *arXiv preprint arXiv:2302.06590*.
Pritchard, Robert D. 1995. *Productivity measurement and improvement: Organizational case studies* (Praeger Publishers/Greenwood Publishing Group).
Puryear, Ben, and Gina Sprint. 2022. 'Github copilot in the classroom: learning to code with AI assistance', *Journal of Computing Sciences in Colleges*, 38: 37-47.
Rossettini, Giacomo, Lia Rodeghiero, Federica Corradi, Chad Cook, Paolo Pillastrini, Andrea Turolla, Greta Castellini, Stefania Chiappinotto, Silvia Gianola, and Alvisa Palese. 2024. 'Comparative accuracy of ChatGPT-4, Microsoft Copilot and Google Gemini in the Italian entrance test for healthcare sciences degrees: a cross-sectional study', *BMC Medical Education*, 24: 694.
Russo, Daniel. 2024. 'Navigating the complexity of generative ai adoption in software engineering', *ACM Transactions on Software Engineering and Methodology*.
Schwartz, Tony. 2010. 'The productivity paradox', *Harvard Business Review*.
Sellen, Abigail, and Eric Horvitz. 2024. 'The rise of the AI Co-Pilot: Lessons for design from aviation and beyond', *Communications of the ACM*, 67: 18-23.

Smit, Danie, Hanlie Smuts, Paul Louw, Julia Pielmeier, and Christina Eidelloth. 2024. 'The impact of GitHub Copilot on developer productivity from a software engineering body of knowledge perspective'.
Sun, Zhiyao, Shuai Che, and Jie Wang. 2024. 'Deconstruct artificial intelligence's productivity impact: A new technological insight', *Technology in Society*: 102752.
Tepe, Murat, and Emre Emekli. 2024. 'Decoding medical jargon: The use of AI language models (ChatGPT-4, BARD, microsoft copilot) in radiology reports', *Patient Education and Counseling*, 126: 108307.
Toner-Rodgers, Aidan. 2024. 'Artificial Intelligence, Scientific Discovery, and Product Innovation'.
Wermelinger, Michel. 2023. "Using github copilot to solve simple programming problems." In *Proceedings of the 54th ACM Technical Symposium on Computer Science Education V. 1*, 172-78.
Ziegler, Albert, Eirini Kalliamvakou, X Alice Li, Andrew Rice, Devon Rifkin, Shawn Simister, Ganesh Sittampalam, and Edward Aftandilian. 2024. 'Measuring GitHub Copilot's Impact on Productivity', *Communications of the ACM*, 67: 54-63.

## Appendix A – Interview Questionnaire

1. What were your initial expectations regarding M365 Copilot?
2. Do you feel any different now? Please give a specific example of what made your perception change.
3. How do you feel about your productivity and efficiency after using M365 Copilot?
4. Are there specific tasks where the impact of M365 Copilot on productivity and efficiency was more noticeable?
5. Which Microsoft product was most beneficial when used with M365 Copilot?
6. Which Microsoft product was not very useful when used with M365 Copilot?
7. How has M365 Copilot influenced the way you approach and complete your daily tasks?
8. Can you share examples where M365 Copilot contributed to innovative problem-solving?
9. Are there specific types of tasks where M365 Copilot proves more effective?
10. Have you encountered tasks where M365 Copilot's assistance was limited or not useful?
11. How would you describe your overall satisfaction with M365 Copilot?
12. Can you share any positive or negative experiences that shaped your perception or satisfaction with M365 Copilot?
13. How do you perceive the measures taken to ensure data privacy and security in M365 Copilot?
14. Have there been any concerns or incidents related to data ethics during the trial? If yes, can you elaborate on the experience that caused the concerns?
15. Have you noticed any biases in M365 Copilot's functioning or the output it generated?
16. How transparent do you find M365 Copilot in terms of explaining its suggestions?
17. In cases where M365 Copilot makes an error, how easy is it to identify and correct these errors?
18. Who do you think should be held accountable if M365 Copilot's suggestions lead to a mistake or problem in your work?
19. Were you adequately informed about how your data would be used by M365 Copilot before you started using it?
20. Do you feel you have sufficient control over what data M365 Copilot can access and how it is used?
21. Do you believe M365 Copilot could affect employment opportunities within your field by automating tasks traditionally done by humans?
22. Has using M365 Copilot affected the way you develop or apply your professional skills?
23. Have you felt an increased dependence on technology as a result of using M365 Copilot? If so, in what ways?
24. What are your views on the balance between human decision-making and AI recommendations in your workflow?
25. In your opinion, are there specific safeguards or ethical guidelines that should be implemented to govern the use of AI technologies like M365 Copilot in your field?
26. Have you ever felt uncomfortable with the recommendations provided by M365 Copilot due to ethical concerns?
27. Do you think M365 Copilot handles diverse languages and cultural contexts effectively?
28. Have you observed any shortcomings in M365 Copilot that could impact inclusivity, such as biases against certain dialects or cultural references?
29. What long-term ethical concerns might arise from the widespread use of AI technologies like M365 Copilot in professional environments?
30. How should organisations prepare to address the ethical implications of integrating AI more deeply into daily operations?
31. Are there any other ethical issues that you observed while using M365 Copilot?
32. Do you have any other comments regarding your experience with M365 Copilot?